\def\papertitle{Assisted Sound Sample Generation with Musical Conditioning in Adversarial Auto-Encoders}
\def\paperauthorA{Adrien Bitton, Philippe Esling, Antoine Caillon, Martin Fouilleul}
\def\paperauthorB{Author Two}
\def\paperauthorC{Author Three}
\def\paperauthorD{Author Four}

\documentclass[twoside,a4paper]{article}
\usepackage{dafx_19}
\usepackage{amsmath,amssymb,amsfonts,amsthm}
\usepackage{euscript}
\usepackage[latin1]{inputenc}
\usepackage[T1]{fontenc}
\usepackage{ifpdf}

\usepackage[english]{babel}
\usepackage{caption}
\usepackage{subfig} % or can use subcaption package
\usepackage{color}

\usepackage{adjustbox}

\ninept

\usepackage{times}
\newif\ifpdf
\ifx\pdfoutput\relax
\else
   \ifcase\pdfoutput
      \pdffalse
   \else
      \pdftrue
\fi

\ifpdf % compiling with pdflatex
  \usepackage[pdftex,
    pdftitle={\papertitle},
    pdfauthor={\paperauthorA, \paperauthorB, \paperauthorC, \paperauthorD},
    colorlinks=false, % links are activated as colror boxes instead of color text
    bookmarksnumbered, % use section numbers with bookmarks
    pdfstartview=XYZ % start with zoom=100% instead of full screen; especially useful if working with a big screen :-)
  ]{hyperref}
\else % compiling with latex
  \usepackage[dvips]{epsfig,graphicx}
  \usepackage[dvips,
    colorlinks=false, % no color links
    bookmarksnumbered, % use section numbers with bookmarks
    pdfstartview=XYZ % start with zoom=100% instead of full screen
  ]{hyperref}
\fi
  \usepackage[hypcap=true]{caption}
\title{\papertitle}

\affiliation{
\paperauthorA}
{{Institut de Recherche et Coordination Acoustique-Musique (IRCAM)} \\ CNRS - UMR 9912, UPMC - Sorbonne Universite \\ 1 Place Igor Stravinsky, F-75004 Paris, France \\ {\tt adrien.bitton@ircam.fr}}

\begin{document}
% more pdf-tex settings:
\ifpdf % used graphic file format for pdflatex
  \DeclareGraphicsExtensions{.png,.jpg,.pdf}
\else  % used graphic file format for latex
  \DeclareGraphicsExtensions{.eps}
\fi

\maketitle

\begin{abstract}
\textit{Deep generative neural networks} have thrived in the field of computer vision, enabling unprecedented intelligent image processes. Yet the results in audio remain less advanced and many applications are still to be investigated. Our project targets real-time sound synthesis from a reduced set of high-level parameters, including semantic controls that can be adapted to different sound libraries and specific tags. These generative variables should allow expressive modulations of target musical qualities and continuously mix into new styles.
\\
To this extent we train \textit{auto-encoders} on an orchestral database of individual note samples, along with their intrinsic attributes: note class, \textit{timbre domain} (an instrument subset) and \textit{extended playing techniques}. We condition the decoder for explicit control over the rendered note attributes and use \textit{latent adversarial training} for learning expressive style parameters that can ultimately be mixed. We evaluate both generative performances and correlations of the attributes with the latent representation. Our ablation study demonstrates the effectiveness of the \textit{musical conditioning}. 
\\
The proposed model generates individual notes as magnitude spectrograms from any probabilistic latent code samples (each latent point maps to a single note), with expressive control of orchestral timbres and playing styles. Its training data subsets can directly be visualized in the 3-dimensional latent representation. Waveform rendering can be done offline with the \textit{Griffin-Lim algorithm}. In order to allow real-time interactions, we fine-tune the decoder with a pretrained magnitude spectrogram inversion network and embed the full waveform generation pipeline in a \textit{plugin}. Moreover the encoder could be used to process new input samples, after manipulating their latent attribute representation, the decoder can generate sample variations as an \textit{audio effect} would. Our solution remains rather light-weight and fast to train, it can directly be applied to other sound domains, including an \textit{user's libraries} with \textit{custom sound tags} that could be mapped to specific generative controls. As a result, it fosters creativity and intuitive audio style experimentations. Sound examples and additional visualizations are available on Github\footnote{\href{https://github.com/acids-ircam/Expressive_WAE_FADER}{https://github.com/acids-ircam/Expressive\_WAE\_FADER}}, as well as codes after the review process.
\end{abstract}

\section{Introduction}

Modern music production techniques rely on large and heterogeneous sound sample libraries along with diverse digital instruments and effects. It opens to a great variety of sound design possibilities and limitless contents to compose with, however principled interactions and scaled visualisations are still lacking in order to efficiently explore such potential and use it to generate \textit{target sound qualities}.
\\
\textit{Unsupervised generative models} learn an underlying data distribution solely based on the observation of examples, in order to consistently generate novel content. They have been successfully applied to complex computer vision tasks such as processing facial expressions, landscapes, visual styles and paintings. Some solutions to audio emerged more recently, including pioneer musical systems such as \textit{NSynth} (Neural Synthesizer \cite{nsynth}) for real-time high-quality sound synthesis. However, the heavy model architecture and prohibitive training time restrict its dissemination. The learned internal representation remains mostly uninformative and its many generative parameters are still too little correlated to explicit semantic qualities.
\\
In this paper, we develop a high-level sound synthesis system with meaningful data visualisations and explicit musical controls. It is a lighter \textit{non-autoregressive model} that can be trained fast on small datasets, including an user's personal libraries. Our goal is to learn expressive style variables from any sound tags, so that the model fosters creativity and \textit{assists digital interactions} in music production. Considering note samples of orchestral instruments, we could for instance synthesise novel timbres or \textit{playing style hybrids}.
\begin{figure}[h]
\centering
\includegraphics[width=0.4\textwidth\center]{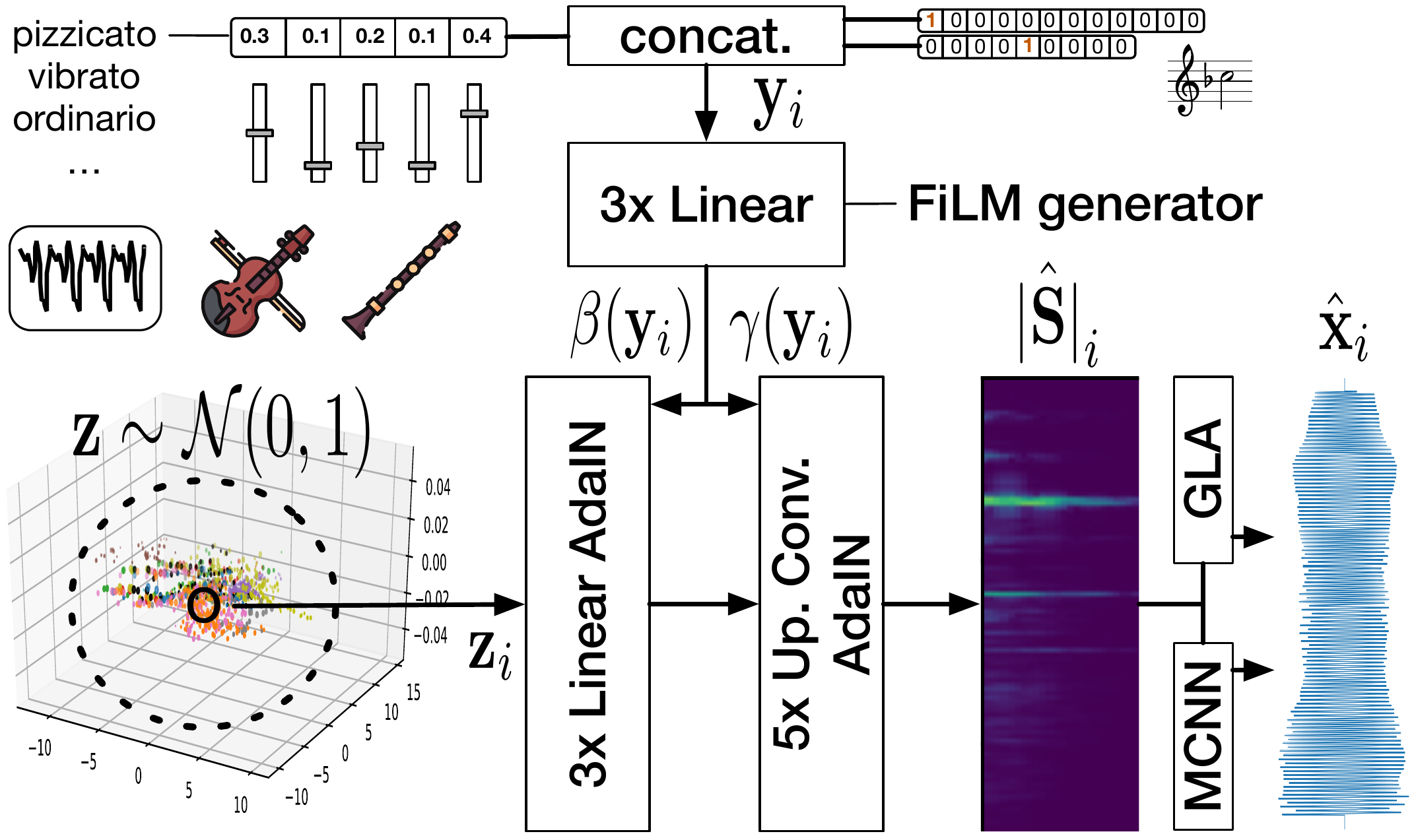}
\caption{High-level note sample generation from the latent representation and musical conditioning in the decoder with FiLM. Intermediate features are modulated by the note targets and expressive style controls in order to synthesize new timbres and effects.}
\end{figure}
\\
We train \textit{Wasserstein Auto-Encoders} (WAEs \cite{WAE}\cite{infoVAE}) on Mel- spectrogram magnitudes to organise a generative latent representation of individual note samples spanning the \textit{tessitura} of 12 orchestral instruments. The considered database has intrinsic attributes: note classes, playing styles and timbres (each instrument subset), that we wish to control when generating new notes from the latent space. Thus we extend the WAE model with musical conditioning in the decoder and \textit{Adaptive Instance Normalization} (AdaIN \cite{AdaIN}). Using \textit{Feature Wise Linear Modulation} (FiLM \cite{FiLM}) and adversarial training with a \textit{Fader latent discriminator} \cite{fader}, our WAE-Fader model effectively learns these generative controls along with expressive style variables that can be mixed continuously.
\\
We evaluate these features in terms of generative performances and representation. We perform an \textit{ablation study} and show that the model can sustain a good test reconstruction quality while achieving an accurate attribute-conditional generation. The success of the method relies on an attribute-free latent representation so that the decoder is pushed to learn the conditioning. These distributions can be visualized directly in the 3-dimensional latent space where clusters denote an undesired attribute encoding. We measure it with inter-class statistics and latent post-classification. The experiment validates correlations between low attribute encoding and effective conditioning.
\\
We obtain an expressive note sample generator with \textit{3-dimensional} representations of the training sound domains, decoding \textit{probabilistic latent samples} with explicit control over the rendered note qualities. The learned style variables of the orchestra can ultimately be mixed continuously, as \textit{faders} do, in order to intuitively explore new musical effects. Generated spectrogram magnitudes can approximately be inverted to waveform with the \textit{Griffin-Lim} iterative algorithm (GLA \cite{GLA}). Ultimately we fine-tune the decoder with a pretrained inversion network \cite{MCNN} for \textit{real-time waveform synthesis}. We embed the resulting generative system in a \textit{plugin} allowing for \textit{MIDI} mapping, live exploration and \textit{Digital Audio Workstation} (DAW) integration.

\section{State-of-art}

\subsection{Generative models and regularized auto-encoders}

\textit{Generative models} aim to find the underlying probability distribution of the data $p(\mathbf{x})$ based on a set of examples in $\mathbf{x}\in\mathbb{R}^{d_{x}}$. To do so, we consider \emph{latent variables} defined in a lower-dimensional space $\mathbf{z}\in\mathbb{R}^{d_{z}}$ ($d_{z} \ll d_{x}$), a higher-level representation that could have led to generate any given example. The latent variable generative model is defined by the joint probability distribution $p(\mathbf{x}, \mathbf{z}) = p(\mathbf{x} \vert \mathbf{z})p(\mathbf{z})$, where the \textit{prior} $p(\mathbf{z})$ is usually modelled with simpler distributions such as Gaussian or uniform while a complex conditional distribution $p(\mathbf{x} \vert \mathbf{z})$ maps latent codes to the data space. The model could be evaluated with the maximum marginal likelihood over the considered dataset. However for complex distributions that could model real-world data, integration cannot be computed in closed form.
\\
Regularized auto-encoders have been used to reformulate the problem as an optimization by jointly learning the generative mapping $p_\theta (\mathbf{x|z}) \in \mathcal G$ and an encoding distribution $q_{\phi}(\mathbf{z}\vert\mathbf{x})\in\mathcal{Q}$ from families $\mathcal G,\mathcal{Q}$ of \textit{approximate densities} both parameterized with neural networks. This was initially proposed through \textit{Variational Inference} in the \textit{Variational Auto-Encoder} (VAE \cite{VAE}) that maximizes a lower bound of the data log-likelihood:
\begin{equation}
\resizebox{.9\hsize}{!}{$
\mathbb{E}_{q_\phi(\mathbf{z|x})} \big[ \log{ p_\theta (\mathbf{x|z}) } \big] - D_{KL} \big[ q_\phi(\mathbf{z|x}) \parallel p_\theta(\mathbf{z}) \big] \leq \log{ p_\theta (\mathbf{x})}
$}
\end{equation}
This amounts to optimizing the \textit{Evidence Lower Bound Objective} (ELBO) that can be interpreted as follow, the first term is the Negative Log-Likelihood (NLL) data reconstruction cost and the second is the Kullback-Leibler Divergence (KLD) that quantifies the error made by using the approximate $q_\phi(\mathbf{z|x})$ rather than the true $p_\theta(\mathbf{z})$. This latent regularization pushes the encoder to remain close to the prior latent density and can be weighted with a $\beta$ parameter that balances these two objectives.
\begin{equation}
\resizebox{.9\hsize}{!}{$
\mathcal{L}_{\theta, \phi}^{\text{ELBO}} = - \mathbb{E}_{q_\phi(\mathbf{z|x})} \big[ \log{ p_\theta (\mathbf{x|z}) } \big] + \beta \cdot D_{KL} \big[ q_\phi(\mathbf{z|x}) \parallel p_\theta(\mathbf{z}) \big]
$}
\end{equation}
The VAE is implemented with a \textit{stochastic encoder} that parameterises an isotropic Gaussian latent distribution $q_\phi(\mathbf{z|x})$\\$\sim \mathcal{N}(\mu_\phi(\mathbf{x}),\sigma_\phi(\mathbf{x}))$ regularized against an unit variance prior. These assumptions allow analytical KLD computation and differentiable latent sampling for direct optimization of the ELBO.
\\
The KLD forces each \textit{individual} latent code to resemble the prior, which implicitly matches the whole encoded distribution. However a fitted ELBO value does not always result in an \textit{effective inference}. Since the latent codes of different inputs are individually regularized, the KLD may prevent the encoder from learning any useful features (\textit{posterior collapse} \cite{VAEfail}) while the decoder only produces $p_\theta (\mathbf{x})$ regardless of the encoded information. Other conflicting solutions of the ELBO lead to undesired solutions and known limitations of VAEs such as \textit{blurriness} of generated samples or \textit{uninformative latent dimensions} (\cite{VAEcoll}).
\\
With justifications stemming both from \textit{Likelihood-free Optimization} (InfoVAE \cite{infoVAE}) and the theory of \textit{Optimal Transport} (WAE \cite{WAE}), a more general framework for training regularized auto-encoders was recently proposed and that we call Wasserstein Auto-Encoders (WAEs). Considering a \textit{deterministic decoder} $G_\theta:\mathbf{z}\rightarrow \mathbf{x}$ and any family of conditional encoder distribution $Q_\phi(\mathbf{z|x})\in\mathcal{Q}$, it is sufficient that the marginal $Q_Z(\mathbf{z}) := \mathbb{E}_X\big[Q(\mathbf{z|x})\big]$ matches any prior $P_Z$. In comparisons with VAEs, WAEs can optimize any non-negative cost function $C$ and \textit{any divergence measure} $D_Z$ between latent distributions, without requiring a stochastic encoder nor restricting the latent model to Gaussian prior:
\begin{equation}
\resizebox{.9\hsize}{!}{$
\mathcal{L}_\text{WAE} := \inf_{Q(\mathbf{z|x})\in\mathcal{Q}} \mathbb{E}_X \mathbb{E}_{Q(\mathbf{z|x})}\big[C(\mathbf{x},G(\mathbf{z}))\big]+\beta\cdot D_Z(Q_Z(\mathbf{z}),P_Z)
$}
\end{equation}
Thus we set our experiment in the more flexible WAE framework. These regularized auto-encoders are powerful unsupervised representation learning models, rather light-weight and fast to train, performing both inference (encoder) and generation (decoder). They are effective on small datasets (hundreds of training examples), learning a structured latent representation with disentangling capacities encouraged when $\beta > 1$. Once trained, probabilistic samples of the latent prior are consistently decoded into new samples and latent interpolations map to smooth data variations.

\subsection{Maximum Mean Discrepancy Regularization}

As shown for VAEs, the choice of \textit{latent divergence} heavily impacts the resulting model performances. Since the point-wise KLD has strong intrinsic limitations, a more flexible regularization is required for WAEs. Such \textit{differentiable} divergence on latent distributions was developed in the \textit{Reproducing Kernel Hilbert Space} (RKHS) as a distance between probabilistic moments $\mu_{p,q}$ computed with a non-parametric kernel $k$:
\begin{align}
|| \mu_p - \mu_q||^2_H  &= \langle \mu_p - \mu_q, \mu_p - \mu_q \rangle \\
&= \mathbb{E}_{p,p}k(x,x') + \mathbb{E}_{q,q}k(y,y') - 2\mathbb{E}_{p,q}k(x,y)\nonumber
\end{align}
It defines the \textit{Maximum Mean Discrepancy} (MMD \cite{MMD}) between two distributions $x\sim p(x)$ and $y\sim q(y)$, where $\mathbb{E}_{p,q}$ is the expectation that can be evaluated with the \textit{Radial Basic Function} (RBF) kernel of free parameter $\Sigma$:
\begin{align}
k_{\text{(RBF)}} (x,y) &= \exp \left( \frac{ ||x - y||^2}{- 2 \Sigma^2} \right) \qquad \qquad
\end{align}
To the extent of latent regularization, MMD can be computed between every deterministic mini-batch encoding $\mathbf{z}_\text{encoder} = Q(\mathbf{x})$ and random samples from any latent prior $\mathbf{z}_\text{prior} \sim P_Z$. Throughout the model optimization, MMD is thus matching the aggregated encoder posterior to the prior rather than regularizing each latent point individually. In comparisons with KLD, WAE-MMD allows for less constrained inference and richer latent representations. For instance, increasing $\beta$ to two orders of magnitude above the reconstruction cost does not impede the decoder training. Since the WAE objective does not optimize the bounded NLL, the overall generative performances can be improved.
\\
Other kernel functions can be used, which may be more discriminating at the expense of heavier computations. Alternatively to MMD, the WAE-GAN uses an adversarial latent discriminator to assess the divergence, thus optimizing a parametric function that could match even closer the encoder to the prior. However, since we consider a low-dimensional latent space of only 3 dimensions and remain with a simple isotropic Gaussian prior, MMD-RBF is sufficient yet light and stable to train on.

\subsection{Conditioning and feature normalizations}

Regularization in auto-encoders encourages disentanglement of independent generative factors onto separate latent dimensions that would in turn control the corresponding decoded data variations. However this is only partly achieved on toy datasets ($\beta$-VAE \cite{betaVAE}) and in most cases the unsupervised latent dimensions are hardly related to explicit generative parameters. An additional supervision signal may be applied to the generative neural network in order to control and render specific attributes of the data. Thus we consider observations $\mathbf{x}$ paired with attribute annotations $\mathbf{y}$, and condition the decoder as $G:\{\mathbf{z,y}\}\rightarrow \mathbf{x}$.
\\
The simplest conditioning for categorical attributes is to encode them into one-hot vectors that are concatenated to the latent codes before being processed by the decoder. However more advanced conditioning techniques have been developed as for visual style transfer, using full images as conditions (conditional style transfer \cite{styleFiLM}). In the \textit{Feature-wise Linear Modulation} (FiLM \cite{FiLM}) approach, a separate generator learns a mapping from any style inputs to adaptive biases $\beta_\text{FiLM} (\mathbf{y})$ and scales $\gamma_\text{FiLM} (\mathbf{y})$ applied to the conditional network computations. This modulation may be placed anywhere within the architecture and proved to be particularly suited to \textit{Adaptive Instance Normalization} (AdaIN \cite{AdaIN}). Considering the \textit{l-th} hidden layer output activations $\mathbf{h}^l = g^l(\mathbf{h}^{l-1})$ of a generative neural network, the conditional modulation is thus be computed as:
\begin{equation}
\text{AdaIN}(\mathbf{h}^l,\mathbf{y}) = \gamma^l_\text{FiLM} (\mathbf{y})\Bigg[ \frac{\mathbf{h}^l-\mu(\mathbf{h}^l)}{\sigma(\mathbf{h}^l)} \Bigg] + \beta^l_\text{FiLM} (\mathbf{y})
\end{equation}
in which mean and standard deviation \{$\mu$,$\sigma$\} are computed across features, independently for each channel and each sample. In the context of style transfer, it can be interpreted as aligning the mean and variance of the content features with those of the style condition. It is a versatile conditioning technique, requiring little additional computations (particularly when applied channel-wise in convolution layers). It also suits well to handling multiple conditions that may more efficiently be mapped throughout the network rather than arbitrarily concatenated to the input. Thus we will use FiLM and AdaIN for conditioning the decoder on both note and style classes. However, such normalization is not suited to classification tasks since content features are individually normalized. In order to preserve its inference power, we will use \textit{Batch Normalization} (BN) on the encoder's hidden activations.

\subsection{Adversarial latent training}

Adversarial regularization was proposed as an alternative to MMD in the WAE-GAN. For simple low-dimensional latent distributions, the expense of an additional parametric adversarial regularizer is not required. Nonetheless, adversarial latent training remains relevant for expressive conditioning. As detailed in the previous section, adaptive conditioning techniques paired with specific feature normalizations substantially improved feed-forward style transfer. However, in an auto-encoder setting, if the latent space implicitly encodes the attributes of interest, the decoder bypasses the conditioning and does not learn any effective generative controls. This problem was tackled in image generation with the introduction of an adversarial \textit{Fader} latent discriminator $\mathcal{F}$ (Fader Networks \cite{fader}) that competes with the non-conditional encoder in order to prevent correlations between attributes and latent distributions. As for the conditional models, we consider annotated data samples $\{\mathbf{x,y}\}$ and for simplicity, a categorical one-hot representation $\mathbf{y}\in\{0,1\}^n$ with a single $y_i=1$ and its opposite $\Bar{\mathbf{y}}:=\mathbf{1}^n-\mathbf{y}$. Such attribute-free latent representation is implemented in two separate optimization steps, first latent classification of the true attribute $\mathcal{F}:\mathbf{z}\rightarrow \hat{\mathbf{y}} \sim p_\psi(\mathbf{y}|Q(\mathbf{x}))$, then adversarial confusion of the latent classifier at predicting the opposite:
\begin{equation}
\begin{split}
\mathcal{L}^\text{class.}(\psi|\phi) = - \sum_\mathbf{x,y} \log(p_\psi(\mathbf{y}|Q(\mathbf{x})))\\
\mathcal{L}^\text{adv.}(\phi|\psi) = - \sum_\mathbf{x,y} \log(p_\psi(\Bar{\mathbf{y}}|Q(\mathbf{x})))
\end{split}
\end{equation}
As the encoder is pushed to remain invariant to attributes, the decoder is forced to learn the conditioning in order to reconstruct every input samples along with their source attributes. Thus it replaces adversarial training in the high-dimensional pixel space with latent attribute confusion in the low-dimensional latent space in order to efficiently learn style transfer variables. Applied to facial expressions, these \textit{Fader} variables can continuously modulate complex visual features such as gender (female $ \leftrightarrow $ male) or age (younger $ \leftrightarrow $ older). Moreover, in mixing several attributes, one could generate new style qualities.

\subsection{Audio synthesis}

Neural networks can be trained on spectrogram magnitudes (and other spectral features) for audio analysis purpose. It eases the subsequent modelling task, often involving pattern detection, from a pre-processed structured sound representation. However, for generative purpose, an inversion from magnitude to waveform is required since the complex phase information was discarded. It is commonly done offline with GLA \cite{GLA}. Further advances in generative neural networks for audio have targeted raw waveform modelling through specific architecture design. \textit{Wavenet} \cite{wavenet}\cite{nsynth} is amongst them the most popular solution. It uses several stacks of dilated causal convolutions in order to aggregate multiple temporal granularities and structure long-term dependencies, which is challenging at the high audio sample rate. The output is a single auto-regressive sample prediction given all the previous sample context $p(x_t|x_1..x_{t-1})$. It results in high-quality real-time audio synthesis. However this sample level modelling requires long training times, heavy architectures that offer little knowledge over their learned features.
\\
The \textit{Multi-head Convolutional Neural Network} (MCNN \cite{MCNN}), a recent alternative for audio waveform modelling, was designed as a feed-forward real-time magnitude spectrogram inversion system that is not restricted to linear frequency scale. It proved to outperform GLA quality for speech. The use of differentiable GPU-based STFT computations enables a faster optimization onto spectral losses, rather than \textit{auto-regressive} sample predictions:
\begin{equation}
\begin{split}
\mathbf{x}\xrightarrow{|\text{STFT}|}|\mathbf{S}|\xrightarrow{\text{MCNN}}\hat{\mathbf{x}}\xrightarrow{\text{STFT}}\hat{\mathbf{S}} \Longrightarrow \mathcal{L}^\text{MCNN}(\mathbf{S},\hat{\mathbf{S}}) \\
\mathcal{L}^\text{MCNN} = \lambda_0\cdot\mathcal{L}^\text{SC} + \lambda_1\cdot\mathcal{L}^\text{logSC} + \lambda_2\cdot\mathcal{L}^\text{IF} + \lambda_3\cdot\mathcal{L}^\text{WP}\\
\end{split}
\end{equation}
where $|\mathbf{S}|$ can be any spectrogram magnitude (including Mel-scaled frequencies). The model is well tailored to audio with multiple heads of 1-dimensional temporal up-sample convolutions. These heads focus on different spectral components and sum into waveform. It remains light-weight and could be adapted in an end-to-end waveform auto-encoder. Four objectives were originally proposed, using the complex STFT for the \textit{Instantaneous Frequency} (IF) and \textit{Weighted Phase} (WP) losses, that we could not optimize successfully. Hence we will only use the \textit{Spectral Convergence} (SC) and log-scale magnitude (logSC) losses:
\begin{equation*}
\mathcal{L}^\text{SC}(\mathbf{S},\hat{\mathbf{S}}) = \| |\mathbf{S}| - |\hat{\mathbf{S}}| \|_F / \| |\mathbf{S}| \|_F \text{ with }\|.\|_F \text{ the Frobenius norm}
\end{equation*}
\begin{equation}\label{eq9}
\mathcal{L}^\text{logSC}(\mathbf{S},\hat{\mathbf{S}}) =  \| \log(|\mathbf{S}|+\epsilon) - \log(|\hat{\mathbf{S}}|+\epsilon) \|_1
\end{equation}

\section{Method}

Our experiment begins with the WAE-MMD, isotropic unit variance Gaussian prior $\mathbf{z}_{\text{prior}} \sim \mathcal{N}(0,1)$, RBF kernel and BN in both encoder and decoder in order to structure a 3-dimensional generative latent sound representation. Given a magnitude spectrogram $|\mathbf{S}|$ and a corresponding set of annotated attributes $\mathbf{y}$, we are learning $Q:|\mathbf{S}|\rightarrow \mathbf{z}$ and $G:\mathbf{z} \rightarrow |\hat{\mathbf{S}}|$ such as $|\mathbf{S}| \approx |\hat{\mathbf{S}}| = G(Q(|\mathbf{S}|))$ with \textit{Binary Cross-Entropy} (BCE) reconstruction cost:
\begin{equation}
\begin{split}
\mathcal L_{\text{WAE}} = \text{BCE}(|\mathbf{S}|,|\hat{\mathbf{S}}|) + \beta\cdot\text{MMD}_\text{RBF}(\mathbf{z},\mathbf{z}_{\text{prior}})\\
\text{BCE}(x,\hat{x}) = -\big[x\log\hat{x}+ (1-x)\log(1-\hat{x})\big]\text{ ; }|x|<1
\end{split}
\end{equation}
We can sample random codes from the latent prior and consistently decode new magnitude samples, however there is no control on the output features.
\begin{figure}[h]
\centering
\includegraphics[width=0.4\textwidth\center]{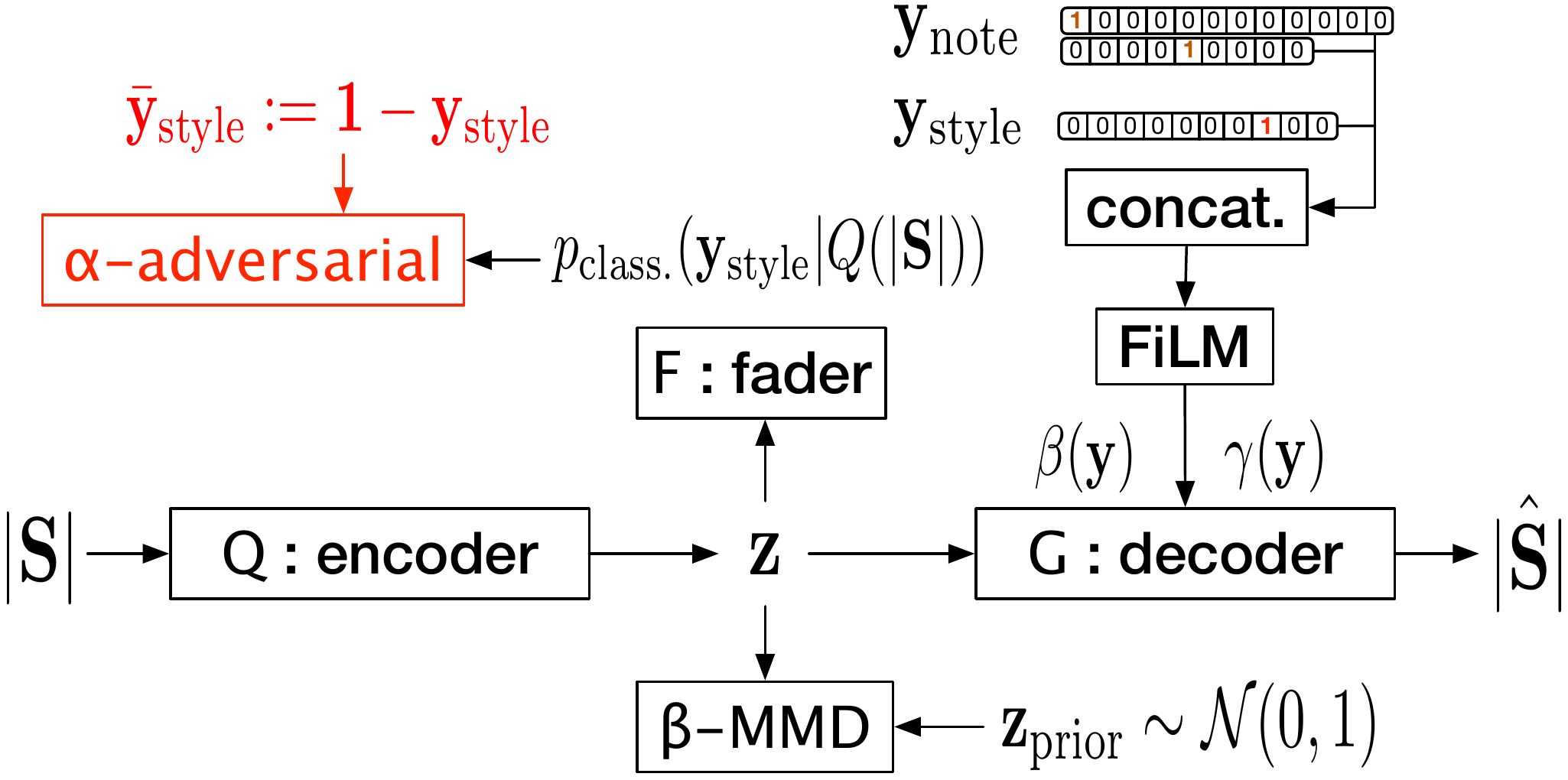}
\caption{How information flows in the adversarial optimization of the WAE-Fader}
\end{figure}
For the orchestra, we consider $\mathbf{y}=\{\mathbf{y}_{\text{note}},\mathbf{y}_{\text{style}}\}$ with $\mathbf{y}_{\text{note}}=\{\text{semitone, octave}\}$. We define the timbre attribute as the class of an instrument subset, which comprises the \textit{Ordinario} mode as well as diverse extended playing techniques such as \textit{Staccato}, \textit{Flatterzunge} or \textit{Pizzicato}. When considering a single subset, we thus aim at controlling the playing techniques of the considered instrument as $\mathbf{y}_{\text{style}}$. When considering multiple instruments, instead we aim at controlling the different timbres, either in \textit{Ordinario} or with mixed playing styles within each instrument subset.
\\
For explicit controls over the rendered attributes, we condition the decoder as $G:\{\mathbf{z,y}\} \rightarrow |\hat{\mathbf{S}}|$ using AdaIN. An additional FiLM generator is fed with concatenated one-hot vectors of the three attribute classes (semitone, octave and style). It learns an adaptive mapping to biases $\beta_\text{FiLM} (\mathbf{y})$ and scales $\gamma_\text{FiLM} (\mathbf{y})$ that are used to modulate the normalized decoder activations. In order to effectively learn the style conditioning and expressively modulate timbres or playing techniques, we use adversarial training with a Fader latent discriminator $\mathcal{F}:\mathbf{z}\rightarrow\hat{\mathbf{y}}_{\text{style}}\sim p_\text{class.}(\mathbf{y}_{\text{style}}|Q(|\mathbf{S}|))$ that competes with the non-conditional encoder in classifying the considered styles from latent codes:
\begin{equation}
\begin{split}
\mathcal L_{\text{class.}} = -\log p_\text{class.}(\mathbf{y}_{\text{style}}|Q(|\mathbf{S}|))\\
\mathcal L_{\text{WAE-Fader}} = \mathcal L_{\text{WAE}} -\alpha\cdot\log p_\text{class.}(\bar{\mathbf{y}}_{\text{style}}|Q(|\mathbf{S}|))
\end{split}
\end{equation}
with $\bar{\mathbf{y}}_{\text{style}}:=\textbf{1}-\mathbf{y}_{\text{style}}$ and $\alpha$ that weights the adversarial loss in the encoder. Classification is optimized on the NLL with \textit{Softmax} probilities. The resulting attribute confusion prevents the latent space from implicitly encoding the style distributions, thus the decoder is forced to use the conditioning to reconstruct the source features from the attribute-free code. Ultimately these learned style variables can continuously be mixed, as actual \textit{faders} do. We refer to this final model as WAE-Fader, that still uses MMD regularization. It allows for controlling the strength of each rendered attribute and intuitively exploring hybrid sound effects from any custom tags, here either chosen from extended playing techniques or from diverse orchestral timbre domains.
\\
The resulting generative system maps any latent coordinate $\mathbf{z}\sim\mathcal{N}(0,1)\in\mathbb{R}^3$ to target note spectrograms with expressive musical style controls. Inversion from spectrogram magnitudes to audio waveforms can be done offline with GLA. Alternatively, we pretrain a MCNN on a larger corpus of musical note samples to allow real-time rendering. In order to improve the final audio quality, we fine-tune the full generative model by freezing the encoder parameters and jointly optimizing the learned decoder with the pretrained MCNN as:
\begin{equation}\label{eq12}
\resizebox{.89\hsize}{!}{$
\mathbf{x}\xrightarrow{\text{STFT}}\mathbf{S}\xrightarrow{|\text{Mels}|}|\mathbf{S}|\xrightarrow{Q}\mathbf{z}\xrightarrow{G\circ\text{MCNN}}\hat{\mathbf{x}}\xrightarrow{\text{STFT}}\hat{\mathbf{S}} \Longrightarrow \mathcal{L}^\text{MCNN}(\mathbf{S},\hat{\mathbf{S}})
$}
\end{equation}
This waveform pipeline $\{G\circ\text{MCNN}\}$ is embed in a plugin for live interactions and DAW integration. Using a MIDI interface, we can for instance trigger target note classes $\mathbf{y}_{\text{note}}$ with keys and map the continuous generative parameters to faders. These are the latent dimensions $\mathbf{z}$, that can also be randomly sampled, and most interestingly the adversarially learned style variables $\mathbf{y}_{\text{style}}$ that can be mixed to explore new sound effects.

\section{Experiment}

\subsection{Dataset}

We use the Studio-On-Line (SOL \cite{SOL}) library of around 15000 individual note samples, across the tessitura of 12 orchestral instruments grouped in 4 families and with many extended playing techniques, that may be specific or shared across instrument families. These are \textit{Wind} (Alto-Saxophone, Bassoon, Clarinet, Flute, Oboe), \textit{Brass} (English-Horn, French-Horn, Tenor-Trombone, Trumpet), \textit{String} (Cello, Violin) and \textit{Keyboard} (Piano). Notes are consistently tagged with the intrinsic attributes of the dataset: note classes (12 semitones across 9 octaves), several dynamics and playing styles of every instrument. We define two style experiments for the orchestra. If training on a single instrument, we aim for expressive synthesis of its playing styles. If training on multiple instruments, we aim for timbre control. Each instrument subset defines a timbre domain, either in \textit{Ordinario} (its common mode) or with all styles mixed.
\\
Audio files are down-sampled to 22050Hz and pre-processed into Mel-spectrograms with a FFT size of 2048, hop size of 256 and 500 bins ranging the full spectrum. As we consider a generator of \textit{individual} notes, we set a common audio length of 34560 samples ($\sim$1.6s) from the attack which amounts to 128 STFT frames. We choose this duration as a trade-off between input and latent dimensionality, limiting the amount of silence after shorter playing modes (eg. \textit{Pizzicato}) while keeping some sustain for longer notes (from which some sustain and decay may have been cropped). Magnitudes are floored to 1e-3 and log-scaled in [0,1] according to the BCE range. Each playing style subset of each instrument is split into 80\% training, 10\% validation and 10\% test notes. In average each instrument has 10 playing styles and 100 to 200 notes for each.

\subsection{Implementation details}

\textbf{Architecture of the WAE-Fader:} Our experiments have been implemented in the \textit{PyTorch} environment and our codes will be shared with this dependency. All convolution layers use 2-d. square kernels, an input zero-padding of half the kernel size and are followed by 2-dimensional feature normalization. All fully-connected linear layers are followed by 1-dimensional feature normalization. The non-linear activation used after every normalization is CELU. The deterministic encoder has 5 convolution layers with [12, 24, 48, 96, 128] output channels, kernel size 5 and stride 2, that down-sample the input spectrograms into 128 output maps that are flattened into an intermediate feature vector of size 8192. It is followed with a bottleneck of 3 linear layers of output sizes [1024, 512, 3] mapping to the latent space. For input Mel-spectrograms of size (500,128), it amounts to a dimensionality reduction of more that 5 orders of magnitude. All normalizations are BN. The decoder mirrors this structure with 3 linear layers of output sizes [512, 1024, 8192]. This vector is then reshaped into 128 maps. To avoid the known \textit{checkerboard artifacts} \cite{deconvolution} of the transposed convolution, we use \textit{nearest neighbor} up-sampling followed with convolution of stride 1. These maps are processed with 4 up-sampling of ratios 3, the last one directly mapping to the input dimensionality of (500,128), and 5 convolutions with [96, 48, 24, 12, 1] output channels and kernel sizes [5, 5, 7, 9, 7]. All normalizations are AdaIN and the decoder output activation is sigmoid, bounded in [0,1] according to the BCE range. The FiLM conditioning is applied feature-wise at the output of the first two linear layers and channel-wise after. It amounts to 3688 modulation weights computed by an additional FiLM generator of 3 linear layers of output sizes [512, 1024, 3688]. Its output is split into biases and scales of sizes [512, 1024, 128, 96, 48, 24, 12]. The Fader latent discriminator has 3 linear layers of output sizes [1024, 1024, $\text{n}_\text{style}$] with \textit{LeakyReLU} activations and a dropout ratio of 0.3, mapping latent codes to probabilities of the $\text{n}_\text{style}$ classes.
\begin{figure}[h]
\centering
\includegraphics[width=0.4\textwidth\center]{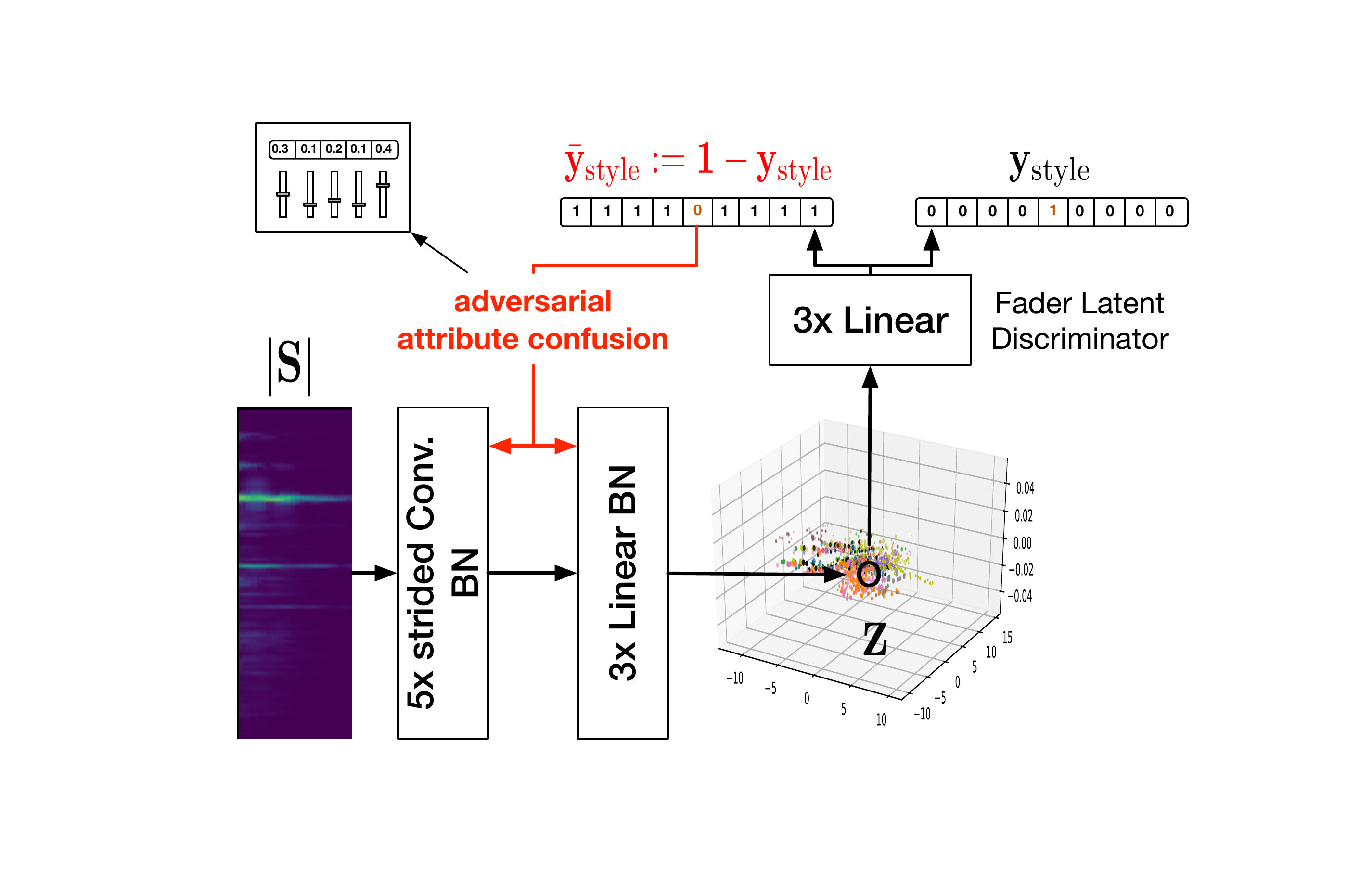}
\caption{The Fader latent discriminator tries to infer the true style attribute while the encoder adversarially aims at fooling it. It encourages attribute invariance in the latent representation and learning of continuous generative controls in the decoder.}
\end{figure}
\\
\\
\textbf{Training parameters:} We train our models with the Adam optimizer, an initial learning rate of 5e-4 and a batch size of 90. All model weights are initialized with Xavier uniform distribution. Depending on the considered data subset size, between 200 and 800 epochs are needed. A single instrument (1000-1500 notes) can be modelled in less than 2 hours on one NVIDIA TITAN Xp GPU. Training over all instruments and styles at once (around 11000 notes) takes less than 12 hours. In the first part of the training (30 to 100 epochs), we only optimize the reconstruction and classification objectives. Then we gradually introduce the MMD regularization ($\beta$-warmup) and the adversarial feedback in the encoder ($\alpha$-warmup) until the first half of training epochs. The rest of the training jointly optimizes all training objectives at their target strengths $\beta$ = 40 and $\alpha$ = 4. These value were estimated in order to approximately balance the gradient magnitudes back-propagated by each loss. However, for the baseline WAE-MMD models we warmup $\beta$ to 500 since it does not prevent from optimizing the reconstruction cost.
\\
\\
\textbf{Signal reconstruction:} The above described model trains on inputs with 128 frames of Mel-spectrogram, which amount to 34560 waveform samples according to our STFT settings. The generated Mel-magnitudes can be approximated back to the linear frequency scale and iteratively inverted with GLA for 100 to 300 iterations. To allow real-time rendering and a possibly improved audio quality, we reproduce the original MCNN architecture for Mel-spectrogram magnitudes inversion. We use 8 heads, $\lambda_0$ = 1 and $\lambda_1$ = 6. We could not successfully optimize the complex losses, however, we compute these magnitude losses on both the linear and Mel frequency scales. We pretrain this model on a larger dataset of around 50 hours audio comprising SOL and subsets of the \textit{Vienna Symphonic Library} (VSL). Ultimately, we fine-tune the trained decoder with this pretrained MCNN. To do so, we freeze the encoder weights and optimize $G\circ\text{MCNN}$ on the model train set. In equation (\ref{eq12}), the auto-encoder pair $G,Q$ maps to Mel-spectrogram magnitudes $|\mathbf{S}|$ which are inverted to signals by the MCNN. However, the loss computation $\mathcal{L}^\text{MCNN}(\mathbf{S},\hat{\mathbf{S}})$ is not necessarily restricted to this frequency scale. Thus we evaluate and sum $\mathcal{L}^\text{SC}$ $\mathcal{L}^\text{logSC}$ from equation (\ref{eq9}) on both linear and Mel frequency scaled magnitudes.

\begin{table}[ht!]
\begin{center}
\begin{tabular}{ |c|c|c|c| } 
\hline
classified attribute ($\text{n}_\text{style}$) & train set & validation & test \\
\hline
\hline
Semitone (12) & 1.00 & 0.99 & 0.99 \\
\hline
Octave (9) & 1.00 & 0.99 & 1.00 \\
\hline
Ordinario timbres (12) & 1.00 & 1.00 & 1.00 \\
\hline
Extended timbres (12) & 1.00 & 1.00 & 1.00 \\
\hline
Violin playing styles (10) & 1.00 & 0.97 & 0.95 \\
\hline
Clarinet playing styles (10) & 1.00 & 0.96 & 0.94 \\
\hline
Piano playing styles (10) & 1.00 & 0.92 & 0.95 \\
\hline
Trumpet playing styles (10) & 1.00 & 0.92 & 0.94 \\
\hline
Alto-Saxophone pl. styles (10) & 1.00 & 0.98 & 1.00 \\
\hline
Tenor-Trombone pl. styles (11) & 1.00 & 0.90 & 0.90 \\
\hline
\end{tabular}
\caption{Reference \textbf{F1-scores} of the pretrained data classifiers used for the evaluation of conditional note generations\label{tab:classif_ref}}
\end{center}
\end{table}

\subsection{Evaluations}

\textbf{Generative performances:} First, we evaluate the ability of our models to produce accurate spectrograms by computing the reconstruction scores on the test set with \textit{Root-Mean Squared Error} (RMSE) and \textit{Log-Spectral Distance} \resizebox{.39\hsize}{!}{$\text{LSD} = \sqrt{\sum \big[10\log_{10}(|\mathbf{S}|/|\hat{\mathbf{S}}|) \big]^2 }$}. Regarding the conditioning aspects, we first pretrain data classifiers to reliably discriminate the different attribute classes and report their performances in Table~\ref{tab:classif_ref}. These classifiers share the same architecture as the encoder but map to the $\text{n}_\text{style}$ classes of interest. We use them as references to evaluate the effectiveness of the conditioning. Then, we sample an evaluation batch of 1000 random latent points from the prior, along with random semitone and octave targets. This evaluation batch is decoded to each attribute of the model (either playing styles or timbres) and classified with the corresponding reference classifier. A high accuracy means an effective conditioning for the task of musical note generation. We report the average accuracy for all the target conditions, with random octaves both in [0-8] (full orchestral range) or in [3-4] where models train on the overlap of every instrument tessitura.
\\
\\
\textbf{Latent space structure:} The effectiveness of the conditioning relies on learning an attribute-free latent representation of the data. If the attribute distributions are clustered, the decoder may learn their correlations with latent dimensions and bypass the conditioning signal. This phenomenon is alleviated with adversarial training of the non-conditional encoder against a Fader latent discriminator. As we map to 3-dimensional spaces, we can directly visualize this latent organization. We also propose two evaluations of the attribute representations. First, we compute the average inter-class latent statistics with MMD. In this case, low values mean that the attribute distributions blend in the final representation. Second, we also perform a post-classification task by training classifiers at predicting the attributes from the learned latent representation. These models use the same architecture as the Fader discriminator, and we report their final accuracy. In this case, low scores mean that the latent representation did not encode the attributes.

\section{Results}

\subsection{Ablation study}

\begin{table}[t]
\begin{center}
\begin{adjustbox}{width=\columnwidth,center}
\begin{tabular}{|c|c|c|c|c|c|c|c|c|}
\hline 
model & \multicolumn{2}{c|}{test rec.} & \multicolumn{4}{c|}{note cond. acc.} & \multicolumn{2}{c|}{style cond. acc.}\tabularnewline
\hline 
 & MSE & LSD & st.$_{34}$ & oct.$_{34}$ & st.$_{08}$ & oct.$_{08}$ & style$_{34}$ & style$_{08}$\tabularnewline
\hline 
\hline 
\multicolumn{9}{|c|}{Violin playing styles (n$_{style}$=10) 1475 training note samples}\tabularnewline
\hline 
WAE-MMD & 0.76 & 68.2 & NA & NA & NA & NA & NA & NA\tabularnewline
\hline 
WAE-note & 0.69 & 55.4 & 0.73 & 0.72 & 0.47 & 0.43 & NA & NA\tabularnewline
\hline 
WAE-style & 0.74 & 59.6 & 0.47 & 0.39 & 0.30 & 0.22 & 0.20 & 0.17\tabularnewline
\hline 
WAE-Fader & 0.80 & 91.1 & \textbf{0.96} & \textbf{0.77} & 0.97 & 0.48 & \textbf{0.88} & \textbf{0.93}\tabularnewline
\hline 
\hline 
\multicolumn{9}{|c|}{Ordinario timbres (n$_{style}$=12) 1784 training note samples}\tabularnewline
\hline 
WAE-MMD & 1.04 & 88.3 & NA & NA & NA & NA & NA & NA\tabularnewline
\hline 
WAE-note & 0.84 & 71.6 & 0.99 & 0.96 & 0.62 & 0.53 & NA & NA\tabularnewline
\hline 
WAE-style & 0.80 & 65.7 & 0.64 & 0.58 & 0.30 & 0.24 & 0.33 & 0.19\tabularnewline
\hline 
WAE-Fader & 1.01 & 105 & \textbf{1.00} & \textbf{1.00} & 0.94 & 0.68 & \textbf{0.95} & \textbf{0.70}\tabularnewline
\hline 
\hline 
\multicolumn{9}{|c|}{Extended timbres (n$_{style}$=12) >11000 training note samples}\tabularnewline
\hline 
WAE-MMD & 0.93 & 175 & NA & NA & NA & NA & NA & NA\tabularnewline
\hline 
WAE-note & 0.69 & 173 & 0.99 & 0.98 & 0.72 & 0.64 & NA & NA\tabularnewline
\hline 
WAE-style & 0.65 & 172 & 0.84 & 0.83 & 0.44 & 0.39 & 0.61 & 0.34\tabularnewline
\hline 
WAE-Fader & 1.32 & 182 & \textbf{1.00} & \textbf{1.00} & 0.90 & 0.71 & \textbf{0.95} & \textbf{0.64}\tabularnewline
\hline 
\end{tabular}
\end{adjustbox}
\caption{The ablation study confirms the effectiveness of the \textbf{WAE-Fader} conditioning, both on target notes and playing styles or timbres. The conditional latent sampling is either performed with random octaves in [3,4] (the overlap of every tessitura) and [0-8] (the full orchestra range), we report the accuracy of the conditioning with respect to the targets $\text{note}_{34,08}$ (st. is semitone classe and oct. is octave classe) and $\text{style}_{34,08}$.\label{tab:abla_gen}}
\end{center}
\end{table}
%according to these notations, we use st. for the semitone classes, oct. for the octave classes.\\Test set spectrogram reconstructions as \textit{test rec.} := [RMSE;LSD].\\Accuracy of the note conditioning as \textit{note cond. acc.} := [$\text{(st.,oct.)}_{34}$ ; $\text{(st.,oct.)}_{08}$], with random octaves in either [3,4] or [0,8].\\Accuracy of the style conditioning as \textit{style cond. acc.} := [$\text{style}_{34}$ ; $\text{style}_{08}$], with random octaves in either [3,4] or [0,8].\\Inter-class latent statistics as \textit{inter-class MMD} := [st.;oct.;styles].\\Post-classification accuracy as \textit{post-class. acc.} := [st.;oct.;styles].
We defined both generative and representation evaluations to assess the effectiveness of our proposed musical conditioning. To study the benefits and compromises of each model feature, we train the base WAE-MMD and compare it with ablations of the WAE-Fader. The incremental model comparisons are WAE-MMD (no conditioning), WAE-note (semitone and octave conditioning), WAE-style (note and style conditioning) and WAE-Fader. In order to simplify the notation, we do not specify the MMD but this regularization is used for all models. We performed this ablation study on the violin subset that has the following annotated playing styles: \textit{Ordinario, Sustained, Short, Non-vibrato, Staccato, Pizzicato-secco, Medium-vibrato-short, Tremolo, Medium-vibrato-sustained} and \textit{Pizzicato-l-vib}. We also compare the WAE-Fader on instrument timbres, either in ordinario or for all extended techniques mixed per instrument subset. Table~\ref{tab:abla_gen} reports the successive generative performances of the models. Table~\ref{tab:abla_lat} reports the latent evaluations, showing how the conditioning tasks are reflected in the learned representations. It confirms the effectiveness of the expressive conditioning when the attribute-invariance assumption is achieved.
\\
As we can see, conditioning WAE-note on the semitone and octave classes shows that the WAE-MMD model can partly learn the note controls with FiLM conditioning. Accordingly, the latent space structure does not exhibit strong correlations with the note classes anymore but with the style attributes that become the main unsupervised data feature. We also notice that this additional supervision improves the reconstruction quality. However, when adding the style conditioning in WAE-style, it seems that most performances drop. Indeed, the overall conditioning becomes little effective, both for the target note and style conditions. The final results show that the adversarial latent training enables the WAE-Fader model to effectively learn the complete conditioning, at the expense of a possible drop in its reconstruction accuracy.
\\
It also seems that the task of modelling the playing styles when learning on a single instrument is more challenging than changing the timbres across multiple instruments. This can be seen in the lower performance of the WAE-style model applied to the violin. This may also be explained by the reduced size of the training data when the learning is restricted to single instrument subsets. These observations are supported by the resulting audio outputs of the conditional note generations. Indeed, it appears that meaningful and expressive variations when switching to any attribute conditions are only achieved with our proposed WAE-Fader model. This is successful for conditioning applied on both timbre attributes or playing styles.
\begin{table}[t]
\begin{center}
\begin{adjustbox}{width=\columnwidth,center}
\begin{tabular}{|c|c|c|c|c|c|c|}
\hline 
model & \multicolumn{3}{c|}{inter-class MMD} & \multicolumn{3}{c|}{post-class. acc.}\tabularnewline
\hline 
 & st. & oct. & style & st. & oct. & style\tabularnewline
\hline 
\hline 
\multicolumn{7}{|c|}{Violin playing styles (n$_{style}$=10) 1475 training note samples}\tabularnewline
\hline 
WAE-MMD & 0.25 & 0.26 & 0.30 & 0.92 & 0.94 & 0.56\tabularnewline
\hline 
WAE-note & 0.03 & 0.10 & 0.50 & 0.04 & 0.49 & 0.82\tabularnewline
\hline 
WAE-style & 0.12 & 0.16 & 0.35 & 0.25 & 0.55 & 0.59\tabularnewline
\hline 
WAE-Fader & 0.02 & 0.01 & 0.46 & 0.08 & 0.34 & 0.64\tabularnewline
\hline 
\hline 
\multicolumn{7}{|c|}{Ordinario timbres (n$_{style}$=12) 1784 training note samples}\tabularnewline
\hline 
WAE-MMD & 0.12 & 0.38 & 0.28 & 0.75 & 0.87 & 0.59\tabularnewline
\hline 
WAE-note & 0.02 & 0.28 & 0.51 & 0.33 & 0.57 & 0.83\tabularnewline
\hline 
WAE-style & 0.04 & 0.40 & 0.35 & 0.13 & 0.60 & 0.63\tabularnewline
\hline 
WAE-Fader & 0.33 & 0.08 & 0.03 & 0.17 & 0.25 & 0.23\tabularnewline
\hline 
\hline 
\multicolumn{7}{|c|}{Extended timbres (n$_{style}$=12) >11000 training note samples}\tabularnewline
\hline 
WAE-MMD & 0.02 & 0.41 & 0.12 & 0.71 & 0.89 & 0.48\tabularnewline
\hline 
WAE-note & 5e-3 & 0.30 & 0.22 & 0.07 & 0.49 & 0.71\tabularnewline
\hline 
WAE-style & 4e-3 & 0.32 & 0.18 & 0.07 & 0.56 & 0.55\tabularnewline
\hline 
WAE-Fader & 3e-3 & 0.20 & 0.11 & 0.11 & 0.46 & 0.43\tabularnewline
\hline 
\end{tabular}
\end{adjustbox}
\caption{The ablation study allows to monitor the latent organization in the different models and throughout their training, as shown in Figure~\ref{fig:lat_plot}. We use both inter-class statistics and latent post-classification to estimate the final attribute invariance in the learned representation.\label{tab:abla_lat}}
\end{center}
\end{table}

\begin{figure}[h]
\centering
\includegraphics[width=0.4\textwidth\center]{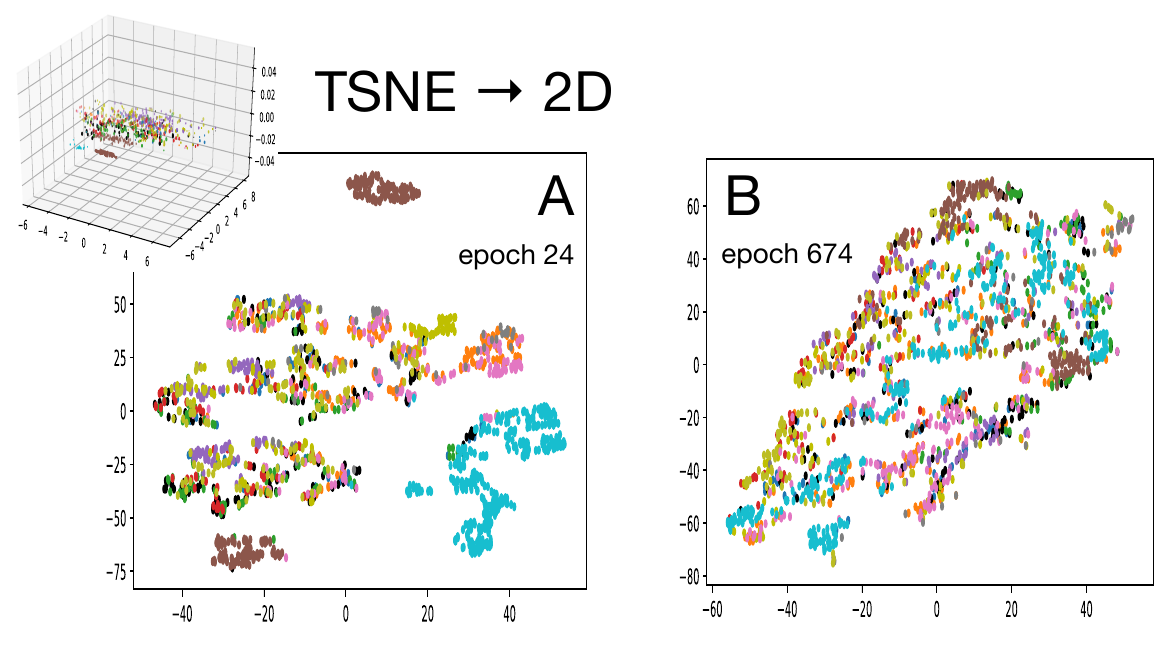}
\caption{Latent organization as the WAE-Fader model trains on the ordinario timbres, each instrument domain being represented by a separate color. In \textbf{\textit{A}}, at epoch 24, the encoder does not optimize the adversarial loss yet. Its unsupervised representation exhibits the attribute classes. In \textbf{\textit{B}}, at epoch 674, the $\alpha$-warmup is finished and the adversarial latent training had blended the attribute distributions. The 2-dimensional projections are performed with t-Distributed Stochastic Neighbor Embedding (TSNE).\label{fig:lat_plot}}
\end{figure}

\subsection{Expressive note sample generations}

\begin{table*}[t]
\begin{center}
\begin{tabular}{|c|c|c|c|c|c|c|c|c|c|c|c|c|c|c|}
\hline 
model & \multicolumn{2}{c|}{test rec.} & \multicolumn{4}{c|}{note cond. acc.} & \multicolumn{2}{c|}{style cond. acc.} & \multicolumn{3}{c|}{inter-class MMD} & \multicolumn{3}{c|}{post-class. acc.}\tabularnewline
\hline 
 & MSE & LSD & st.$_{34}$ & oct.$_{34}$ & st.$_{08}$ & oct.$_{08}$ & style$_{34}$ & style$_{08}$ & st. & oct. & style & st. & oct. & style\tabularnewline
\hline 
\hline 
Clarinet & 0.87 & 116 & 0.96 & 0.99 & 0.98 & 0.45 & 0.97 & 0.92 & 0.05 & 0.58 & 0.12 & 0.15 & 0.76 & 0.41\tabularnewline
\hline 
Piano & 0.99 & 113 & 0.53 & 0.91 & 0.47 & 0.72 & 0.72 & 0.64 & 0.03 & 0.03 & 0.08 & 0.16 & 0.20 & 0.43\tabularnewline
\hline 
Trumpet & 0.90 & 107 & 0.91 & 0.93 & 0.96 & 0.37 & 0.90 & 0.87 & 0.60 & 0.11 & 0.02 & 0.42 & 0.50 & 0.29\tabularnewline
\hline 
Alto-Sax. & 1.22 & 131 & 0.96 & 0.99 & 0.98 & 0.40 & 0.76 & 0.71 & 0.14 & 0.09 & 0.50 & 0.08 & 0.48 & 0.48\tabularnewline
\hline 
T. Trombone & 0.96 & 100 & 1.00 & 1.00 & 0.92 & 0.41 & 0.83 & 0.77 & 0.04 & 0.14 & 0.34 & 0.06 & 0.55 & 0.47\tabularnewline
\hline 
\end{tabular}

\caption{Additional WAE-Fader results on the playing techniques of instruments in other orchestral families\label{tab:addFader}}
\end{center}
\end{table*}
%The results detailed in the ablation study show the WAE-Fader model performances on the 12 considered orchestral timbres as well as on the Violin playing styles. 
In this section, we report additional experiments on the WAE-Fader models when conditioned on the playing styles of different instruments and families. As shown in Table~\ref{tab:addFader}, our model seems to train successfully on playing styles in every instrument families, as well as across the 12 instrument timbres of the orchestra as shown in the previous ablation study. This amounts to a great variety of sound qualities spanning extended modes of the orchestra, and let us hypothesize that the model could be applied to other sound domains as long as the tags are consistent with the data. Furthermore, the style variables learned with the Fader latent discriminator are continuous independent controls that can be mixed. Hence, this can allow our system to modulate the strength of rendered styles and create new effects by combining multiple attributes. Our model can also be used for sample modifications, akin to traditional audio effects, by encoding a given sample and manipulating the attribute conditions in order to decode different sample transformations.

\subsection{Audio outputs and plugin development}

As discussed previously, our proposed models can generate magnitude spectrograms, while controling their expressive qualities. These spectrograms can be either inverted to waveform offline with GLA or real-time if paired with MCNN. When fine-tuning the learned decoders with the pretrained MCNN on magnitude losses, we obtain a quality almost equivalent to the GLA approximation. We provide audio examples of test set reconstructions and conditional note generations inverted with both GLA and MCNN for individual listening evaluation on the companion webpage. While the audio quality of these results can still be improved, we can already confirm the ability of the model to provide semantic controls. As the learned style variables of WAE-Fader can be mixed continuously, we also provide some sound examples that were generated when modifying multiple orchestral attributes.
\\
Our proposal provides intuitive sound synthesis of target sound qualities with learned style variables that can be modulated and combined. The unsupervised latent dimensions organize remaining data features, which can be directly visualized in a 3-d. space, in order to perform sampling or explicit control. These features allow to generate timbres, playing styles and hybrid effects across multiple attribute combinations through intuitive interactions. We provide a real-time implementation of our models by relying on the fine-tuned $\{{G\circ\text{MCNN}}\}$ generation. This implementation relies on the \textit{LibTorch} C++ API, which converts trained \textit{PyTorch} models, that we further embed in a \textit{PureData} external. This plugin can be mapped to a MIDI controller or integrated in a DAW for composition and musical performance. This allows to play notes with a keyboard, while using continuous faders to control latent coordinates and mix style conditions.

\section{Conclusion}

We developed an expressive musical conditioning of the Wasserstein Auto-Encoders able to model a collection of orchestral note samples. The model learns effective target semitone and octave controls as well as continuous style variables. We considered extended playing techniques and timbre subsets as attributes, and used adversarial latent training to encourage an attribute-invariant representation in the WAE-Fader. Our ablation study validates the effectiveness of style conditioning when this invariance condition is obtained.
\\
We fine-tuned the decoders with a Mel magnitude spectrogram inversion network that allows real-time waveform rendering and are currently working on refining the audio quality. This results in a note sample generator with meaningful data visualizations and intuitive controls of audio styles. These parameters can be mixed, as faders, in order to explore hybrid sound effects. Our final generative model is embed in a plugin for MIDI mapping and live interactions. This system provides assisted music production and fosters creative sound experimentations. We provide sound examples from our orchestral models, either inverted offline with GLA or with the fine-tuned waveform generation pipeline. These sounds allow for subjective evaluation of both semantic and audio qualities of our solution.
\\
Although we used clearly defined metadata attributes pertaining to instrumental playing styles, the model can potentially be applied to any sound domain. For instance, a user library with custom tags could be mapped to sound synthesis parameters. Furthermore, as the architecture is rather light and scales to small datasets, it could be trained on user libraries. Future experiments will target the quality of the waveform modelling systems for variable note lengths and real-time synthesis. Ultimately, our models could be implemented as a standalone instrument with physical controls that can be mapped to pretrained style variables. This would allow an intuitive and creative exploration across a vast amount of sound variations with a reduced set of adaptive parameters.

\section{Acknowledgements}
This work was supported by the MAKIMOno project 17-CE38-0015-01 funded by the French ANR and the Canadian NSERC (STPG 507004-17), the ACTOR Partnership funded by the Canadian SSHRC (895-2018-1023) and an NVIDIA GPU Grant. We acknowledge and thank the initial contributions of Jean-Baptiste Dakeyo and Geoffroy Thibault, as well as the further collaboration of Antoine Caillon and Martin Fouilleul.

%\newpage
%\nocite{*}
\bibliographystyle{IEEEbib}
\bibliography{main} % requires file DAFx19.bib

\end{document}